\date{}
\documentclass[12pt]{article}

\usepackage{hyperref}
\usepackage[utf8]{inputenc}
\usepackage{amsmath,pgfplots}
\usepackage{authblk}

\usepackage{amssymb}

\title{Solving classification tasks by a receptron based on nonlinear optical speckle fields}

\providecommand{\keywords}[1]
{
  \small	
  \textbf{\textit{Keywords---}} #1
}

\author[1]{B. Paroli}
\author[1]{G. Martini}
\author[1]{M.A.C. Potenza}
\author[1]{M. Siano}
\author[1]{M. Mirigliano}
\author[1]{P. Milani}
\affil[1]{CIMAINA and Dipartimento di Fisica, Università degli Studi di Milano - via G. Celoria, 16, 20133 Milano, Italy}

\begin{document}

\maketitle




\begin{abstract}
Among several approaches to tackle the problem of energy consumption in modern computing systems, two solutions are currently investigated: one consists of artificial neural networks (ANNs) based on photonic technologies, the other is a different paradigm compared to ANNs and it is based on random networks of nonlinear nanoscale junctions resulting from the assembling of nanoparticles or nanowires as substrates for neuromorphic computing. These networks show the presence of emergent complexity and collective phenomena in analogy with biological neural networks characterized by self-organization, redundancy, non-linearity. Starting from this background, we propose and formalize a generalization of the perceptron model to describe a classification device based on a network of interacting units where the input weights are nonlinearly dependent. We show that this model, called “receptron”, provides substantial advantages compared to the perceptron as, for example, the solution of non-linearly separable Boolean functions with a single device. The receptron model is used as a starting point for the implementation of an all-optical device that exploits the non-linearity of optical speckle fields produced by a solid scatterer. By encoding these speckle fields we generated a large variety of target Boolean functions without the need for time-consuming machine learning algorithms. We demonstrate that by properly setting the model parameters, different classes of functions with different multiplicity can be solved efficiently. The optical implementation of the receptron scheme opens the way for the fabrication of a completely new class of optical devices for neuromorphic data processing based on a very simple hardware.
\end{abstract}





\section{Introduction}
\label{sec:introduction}
The current prominent approach to neuromorphic computing is based on the notion that the modification of synaptic connection strengths in the brain allows to learn and to perform complex tasks \cite{Lynn,Lillicrap,Ambrogio,Rajendran,Pershin}. The possibility of adjusting independently the synaptic weights to respond to different input patterns has been translated in the perceptron learning rule where a biological or artificial neuron is modelled as a simple linear summation and thresholding device \cite{Rosenblatt,McCulloch,Jeong, Nagy}. Ensembles of connected perceptrons constitute artificial neural networks (ANNs) where the weight of each unit can be independently modified, accordingly to a learning rule, to obtain a target output, given a certain input \cite{Schuman,Xia,Liu}. The use of crossbar memristive arrays has been proposed in order to reduce the energy payload of ANNs: in this approach all nodes are linearly independent and the equivalent of the corresponding synaptic weight must be updated avoiding cross-talking between individual synapse nodes \cite{Rajendran,Xia,Burr}.

ANNs based on photonic technologies developed for telecom applications \cite{Tomson,Nakajima} can represent a valid alternative to conventional electronic hardware for the achievement of a significant reduction of the operational power and increase of the speed and parallelism \cite{Shen}. Various photonic ANN models have been reported \cite{Nakajima,Shen,Vandoorne}: as in the case of their electronic counterpart, the architecture of optical ANNs is characterized by a lack of similarity with respect to biological neural systems, where self-organization, redundancy, non-linearity, and non-locality governs both structure and functions \cite{Milano_2,Diaz,Hochstetter}. Neurons utilize a wealth of nonlinear mechanisms to transform synaptic input into output firing \cite{Lo,Tononi}; the majority of inputs to a neuron is received primarily through synapses made onto elaborate treelike structures called dendrites \cite{Hausser}, allowing transformations that extend far beyond the simple sum-and-threshold operation \cite{Bicknell}. The morphology and the electrical properties of dendrites define the input-output relationship of neurons and the rules for the induction of synaptic plasticity \cite{Silver}.

A radical alternative to the top-down fabrication of electronic or optical ANNs is based on the use of networks consisting of a large number of nonlinear nanoscale junctions resulting from the random assembling of nanoparticles or nanowires \cite{Li}. These networks show the presence of emergent complexity and collective phenomena in analogy with biological neural networks and, in particular, hierarchical collective dynamics \cite{Mallinson}, and heterosynaptic plasticity \cite{Milano}.

Networks of interconnected nanojunctions are characterized by the nonlinear and distributed nature of the junction weight interactions: the weights are not univocally related to a single node since the highly interconnected junctions regulate their connectivity and the topology of conducting pathways depending on the input stimuli \cite{Mirigliano,Martini,Mirigliano_2}, in analogy to what observed in neuronal dendrites \cite{Bicknell}. This aspect has been somehow neglected up to now, not considering that a consequence is the impossibility of applying the perceptron model to describe the evolution of the synaptic weights upon the interaction with external stimuli \cite{Poirazi}.

Recently we have shown that nanostructured Au films fabricated by assembling gold clusters, produced in the gas phase, have complex nonlinear electrical properties and resistive switching behavior \cite{Mirigliano,Martini,Mirigliano_3,Mirigliano_4}. By interconnecting a generic pattern of electrodes with a cluster-assembled Au film, we demonstrated the fabrication of a device that can perform the binary classification of input signals, following a thresholding process, to generate a set of Boolean functions \cite{Mirigliano}. Considering the non-linear conduction properties of cluster-assembled gold films and their non-local response to input signals we underlined the inadequacy of a perceptron model with linearly independent weights proposing a model called “receptron” where the weights are not just associated with each input, but with their combinations\cite{Mirigliano,Martini}.

Here, we provide a general formalization of the receptron showing the fundamental differences between a receptron and a perceptron and we demonstrate that the general receptron model is not confined to electrical networks but it can be used as a starting point for the implementation of an all-optical device performing classification. We use a training approach based on a search procedure that does not require time-consuming iterative algorithms. An experimental demonstration of the use of an optical receptron for classification tasks is reported.

\section{The receptron model}
\label{sec:model}

Starting from the traditional perceptron model \cite{Minsky} based on linearly independent weights: 

\begin{equation}\label{sommaper}
S=\sum_{j=1}^n{x_j w_j^P},
\end{equation} 

where $j$ numbers the inputs ($j \in [1,n]$) and $w_j^P$ are constant real values, we formally introduce a more general form of Eq. \ref{sommaper}, which allows for the non linear interaction  of the inputs,

\begin{equation}\label{somma}
S=\sum_{j=1}^n{x_j\tilde{w_j}(\vec{x})} \: | \: S \in R,
\end{equation} 

where $\tilde{w_j}(\vec{x}):R^n \rightarrow C$ are complex-valued functions and $\vec{x}=(x_1,...,x_n)$ is the input vector.

Equation \ref{somma} is the basis of the receptron model: while Eq. \ref{sommaper} is a linear combination of the inputs (the weights are constant), in the receptron a modification of the inputs leads, in general, to a variation of the weights value $\tilde{w_j}(\vec{x})$, making the system extremely complex and allowing for the solution of problems not solvable through the simpler rules of a linear system. 

As in the perceptron case the summation in Eq. \ref{somma} origins the activation of the receptron output through the thresholding process,

\begin{equation}\label{condizioni}
Y(x_1,...,x_n) =
\left\{
\begin{array}{rl}
1 & \mbox{  } S>th  \\
0 & \mbox{  } S\leq th \:,
\end{array}
\right.
\end{equation}

where $th$ is a constant threshold parameter. Equation \ref{condizioni} can be written by using the Heaviside function as

\begin{equation}\label{condizioni_2}
Y(x_1,...,x_n) =\Theta \left(S'\right),
\end{equation}

where $S'=b+S$ and $b$ is a constant bias.

With the aim of investigating the non-linear and statistical properties as well as the computing performance of the proposed model, we limit the model to purely Boolean inputs ($x_j \in \{0,1\}$), this facilitates the analysis and helps to highlight the most important features. In this case, the weight functions $\tilde{w_j}(\vec{x})$ can be written with a finite number of parameters $w_{j_1\dots j_n}$, simplifying the model representation. In particular, we can Taylor-expand $\tilde{w_j}(\vec{x})$ and use the idempotency of Boolean variables $(x_j)^q=x_j \forall q \geq 1$ such that $S'=b+\sum_{j=1}^n{x_j\tilde{w_j}(\vec{x})}$ can be written as

\begin{equation}\label{somma_primata}
    S'(\vec{x})=b+\sum_j w_jx_j+\sum_{j<k} w_{jk}x_j x_k +\sum_{j<k<l} w_{jkl}x_j x_k x_l+\cdot\cdot\cdot\:,
\end{equation}

where $w_{j_1\dots j_n}$ are independent parameters that can be seen as the components of a tensor $W$ (“weight tensor”) of rank $n$ and type $(n,0)$. In Eq. \ref{somma_primata} we have contracted, for simplicity of notation, the indexes of the matrix elements $w_{j_1\dots j_n}$ to a single index when all the indexes at the right are equals, as shown here below for $n=3$,

\begin{equation}\label{rule_1}
\begin{split}
w_j \implies j=k=l\\
w_{jk} \implies k=l\\
\end{split}
\end{equation}

or more in general

\begin{equation}\label{rule_2}
w_{j_1...j_m} \implies j_m=j_{m+1}=...=j_n\: .
\end{equation}

The expression in equation \ref{somma_primata} is the most general Boolean-input function. In fact, a function defined over a finite set, such as the case of $S'$, is completely specified by the values assumed over all the elements inside the domain. In the case of Boolean inputs, the domain contains $2^n$ elements: tuning every one of them allows to define the value of the function for each input. The number of coefficients in equation \ref{somma_primata} is 1 for the bias, $n$ for the diagonal elements $w_j$, $\frac{n(n-1)}{2}$ for $w_{jk}$ and finally $\frac{n!}{m!(n-m)!}$ for the m-th degree term, thus the total number is given by

\begin{equation}\label{number_coeff}
\sum_{m=0}^{n} \binom{n}{m} = 2^n\: ,
\end{equation}

where we used the binomial theorem.

We have thus proved that the total number of independent parameters equals the number of input combinations, i.e., we have different degrees of freedom for each input, that can be adjusted to implement any desired output.

The sum in Eq. \ref{somma_primata} reduces to the perceptron case when off-diagonal terms of $W$ vanish. As an example we consider $n=2$, where we write the sum \ref{somma_primata} to 

\begin{equation}\label{example}
S'(\vec{x})=b+x_1 w_{11} +x_2 w_{22}+x_1 x_2 w_{12}.
\end{equation}    

In the perceptron case, the vanishing of $w_{12}$ implies linearity, $S(1,1)=S(0,1)+S(1,0)$. On the contrary, the off-diagonal elements yield $S(1,1) \neq S(0,1)+S(1,0)$ for a receptron, meaning the superposition principle is no longer valid, the latter terms being precisely responsible of the more complex non-linear interaction between the inputs. 

\begin{figure}[http!]
\centering
\includegraphics[scale=0.5]{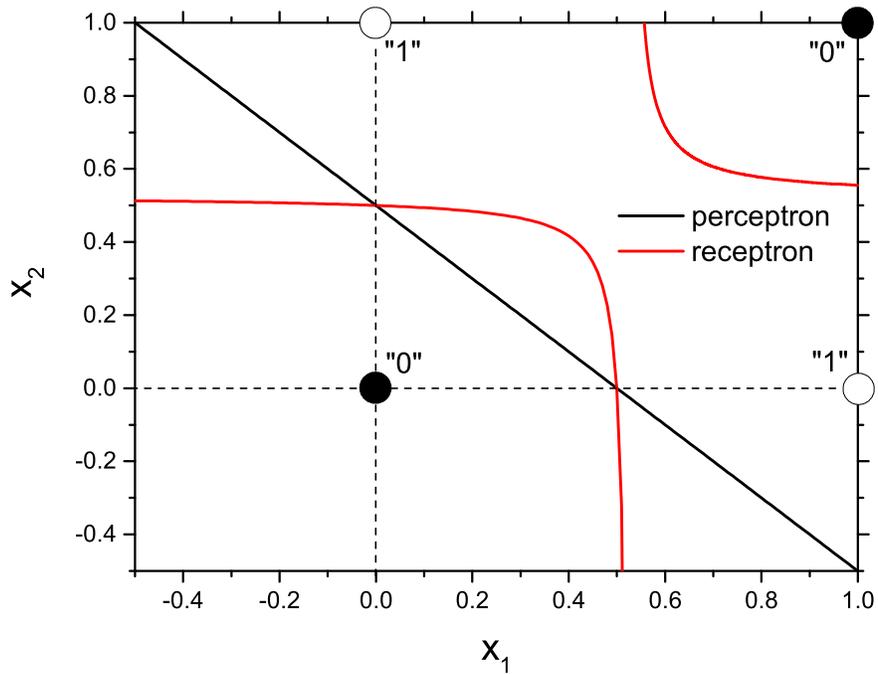}
\caption{\label{xor}Classification of the XOR output states with a single receptron. The black straight line and the red hyperbole are the decision boundaries for the perceptron and a receptron respectively, obtained by posing $S'(\vec{x})=0$ in Eq. \ref{example}. The high states (white circles) are separable from the low states (black circles) with the red hyperbole obtained with the receptron parameters but, as is well known, they cannot be separated with the straight generated by a single  perceptron.}
\end{figure}

Due to the missing coefficients, the number of independent parameters for a perceptron grows proportionally to the number of inputs, while the growth is exponential in a receptron (see Eq. \ref{number_coeff}), highlighting its higher complexity. Therefore, classification of non-linearly separable functions can be realized with a single receptron, as shown in Fig. \ref{xor}. 

Here the output states of the XOR function are shown with black (low state) and white (high state) circles. The black straight line calculated with Eq. \ref{example} by posing $S'(x)=0$ and $w_{12}=0$ cannot separate the black and white circles, therefore the XOR function cannot be implemented with a perceptron, contrarily the non-zero off-diagonal elements transform the straight line in the red hyperbole, separating the black and white circles, and making the XOR easy to classify.

\subsection{Optical receptron}
\label{sec:optical_model}

In this section, we use the model formalized above to describe an optical receptron exploiting the interference of a large number of uncorrelated point-like sources, generating the non-linear interaction between optical inputs discussed above. The development of the optical model allows the explicit analytical calculation of the weights, useful for predicting the overall behaviour of the device in terms of operation and degree of non-linearity.

\begin{figure}[htp]
\centering
\includegraphics[scale=0.6]{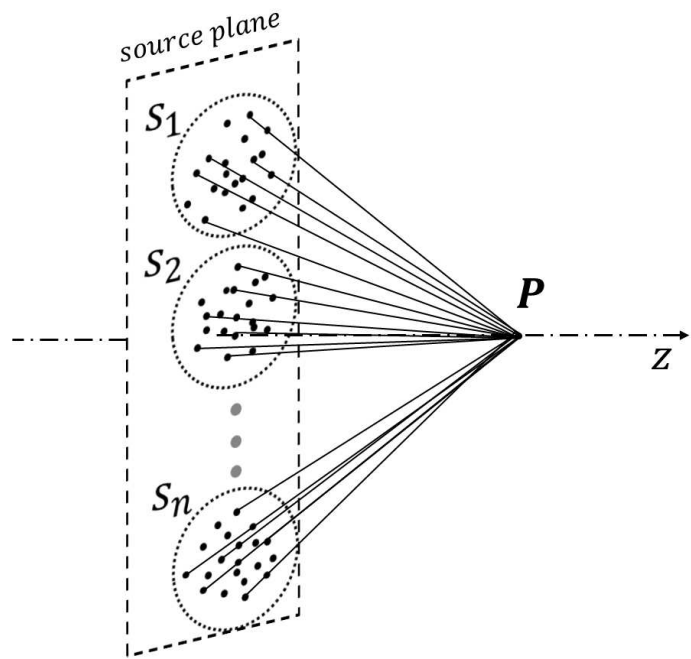}
\caption{\label{optical}Sketch of the optical implementation of a receptron exploiting the interference to create the non-linear interaction $I(\vec{x})=S(\vec{x})$ of  Eq. \ref{reale}. The group of point-like sources $s_j$ are independently on/off modulated through the Boolean inputs $x_j$. The intensity is observed at the point $P$ along the optical axis $z$. }
\end{figure}

Let us consider a set of $n$ sources $s_j$ positioned in a plane as sketched in Fig. \ref{optical}. Since sources are uncorrelated, each will produce an electric field $E_j$ with amplitudes $A_j$ and phases $\phi_j$ at a given observation point $P$ positioned at a distance $d$ from the source plane

\begin{equation}\label{complex_field}
E_j=A_j e^{i\phi_j}\:.
\end{equation}

Let us assume now that the light emission of each group of sources can be turned ON or OFF. This action can be formalized in our model, multiplying the field amplitudes in Eq. \ref{complex_field} with input Boolean variables $x_j\in \{0,1\}$ as

\begin{equation}\label{complex_field_input}
E_j=x_j A_j e^{i\phi_j}.
\end{equation}

The overall intensity in $P$ will be given by

\begin{equation}\label{intensity}
I={\left|\sum_{j=1}^n x_j A_j e^{i\phi_j} \right|}^2 \: .
\end{equation}

If we interpret the intensity at a point on the observation plane as the interaction between inputs (the input being the on/off pattern of sources) we see that the squared modulus in Eq. \ref{intensity} allows for the nonlinearity. We can calculate the weights of the optical receptron to show that this, rather than the perceptron, fits best such a device:

\begin{equation}\label{reale}
\begin{split}
    I&= \sum_{j=1}^n x_j^2 A_j^2 + \sum_{j\neq k}x_j x_k A_j A_k e^{i (\phi_j-\phi_k)}  =\\
    & = \sum_{j=1}^n x_j A_j^2 + \sum_{j<k}x_j x_k A_j A_k e^{i (\phi_j-\phi_k)} + \sum_{j<k}x_j x_k A_j A_k e^{i (\phi_k-\phi_j)} =\\
    & = \sum_{j=1}^n x_j A_j^2 + 2\cdot \sum_{j<k}x_j x_k A_j A_k cos(\phi_j-\phi_k)\: .
\end{split}
\end{equation}

Equation \ref {reale} shows that the intensity produced by the interference of groups of sources as a function of the input Boolean variables is consistent with the summation $S(\vec{x})$ of the proposed receptron model in Eq. \ref{somma}. 

Hence we pose $I(\vec{x})=S(\vec{x})$ to find the weight functions of the optical receptron as

\begin{equation}
\tilde{w_j}(\vec{x})=x_jA_j^2 + \sum_{\substack{k=1 \\ k\neq j}}^Nx_k A_j A_k e^{i (\phi_j-\phi_k)}\:.
\end{equation}

Combining Eq. \ref{reale} and Eq. \ref{somma_primata} we find the components of the weight tensor,

\begin{gather}\label{components}
    w_j = A_j^2\\
    w_{jk} = 2A_j A_k cos(\phi_j - \phi_k)\\
    w_{j_1..j_n}=0\quad elsewhere,
\end{gather}

which demonstrate that the system can be fully described with the proposed model. The large number of sources used in the model is essential to impart high variability of the intensity in the space (depending on many independent parameters): a limited number of sources would still generate non-linearity of the intensity as a function of the inputs $x_j$, but the limited number of free parameters $A_j$, $\phi_j$ would produce a more regular field. Our idea is to maximize the variability of the intensities $A_j$ and phases $\phi_j$ in the different observation points through a speckle field \cite{Speckle,Speckle_1} generated with a random scatterer: this will in turn imply a higher spatial variability of the function $I(\vec{x}) = S(\vec{x})$ implemented by the network. The group of sources described above, is experimentally realized with a scatterer illuminated by means of independent laser beams, as discussed in the next section. 

\section{Experimental setup}
\label{sec:setup}

\begin{figure}[http]
\includegraphics[scale=0.65]{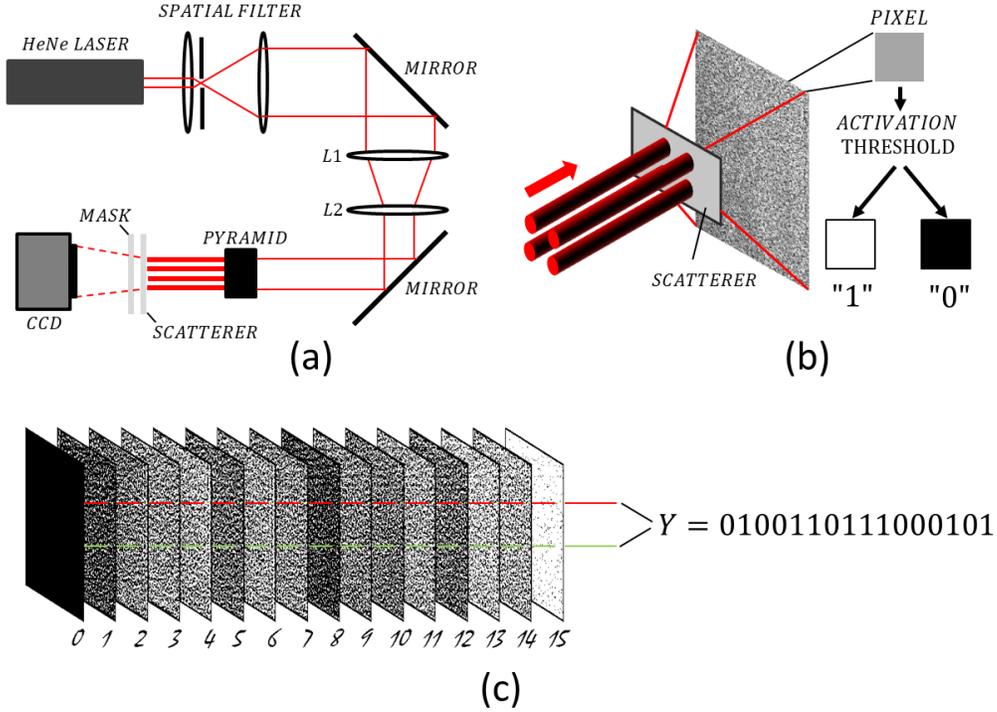}
\caption{\label{setup}(a) Experimental setup for the optical implementation of the receptron. A He-Ne laser beam is spatially filtered and collimated. The positive lens L1 and L2 are used to match the beam size to the optical pyramid base. The pyramid splits the main beam in four independent beams (the receptron inputs) impinging on the scatterer. (b) Sketch of the method used to convert the 8 bit grey scale image of the speckle field in a corresponding black and white Boolean field. The speckle field, shown in grey scale, is perpendicular to the propagation axis of the laser beams (inputs) and generated with the optical pyramid (not shown). The scatterer plane is parallel to the CCD plane. Binarization is realized computationally for all the CCD pixels through a thresholding operation that converts each grey pixel in a black or white pixel as described in Eq. \ref{condizioni}. (c) Example of solution of two identical Boolean functions. A stack of Boolean fields is obtained by changing the 16 input combinations. The two pixels shown in green and red, along the stack, generate the same output sequence $Y=0100110111000101$ as a function of the inputs from 0 to 15. }
\end{figure}

The experimental setup was realized for a 4-input device as shown in Fig. \ref{setup} (a). Here a 5 mW He-Ne laser source was spatially filtered, collimated and split in four independent beams by means of a square optical-glass pyramid (the base of the pyramid is perpendicular to the optical axis). Each beam was sent to a mask comprising four optical apertures which physically implement the $x_j$ of equation \ref{complex_field_input}: all 16 ($2^n$) input combinations can be realized opening and closing the apertures. 

The beams at the exit of the apertures impinge on an optical polypropylene scatterer ($\approx$ 100 $\mu$m thick) at 75 mm from the pyramid, in order to create four independent groups of uncorrelated point like sources as described in section \ref{sec:optical_model}. The scatterer generates a homodyne speckle field that is detected with a Charge Coupled Device (CCD) on an observation plane, orthogonal to the propagation axis of the beams, at a distance of 125 mm from the scatterer. This distance is chosen in such a way the $\approx$ 40 $\mu$m speckle size (FWHM) is slightly larger than the CCD pixel size (7.4 $\mu$m). The speckle field is acquired in a 8 bit grey scale. The activation of each pixel, corresponding to a single receptron output, was implemented computationally by the thresholding operation (as shown in Eq. \ref{condizioni}), that converts the grey scale speckle pattern in a black and white Boolean field (see Fig. \ref{setup} (b)). The high-resolution of the CCD ($1600 \times 1200$ pixels) simultaneously detects a multitude of receptrons (with common inputs) in a completely parallel fashion.

\section{Results and discussion}
\label{sec:results}

\subsection{Uniform speckle fields}
\label{sec:uniformity}
\begin{figure}[http]
\includegraphics[scale=0.6]{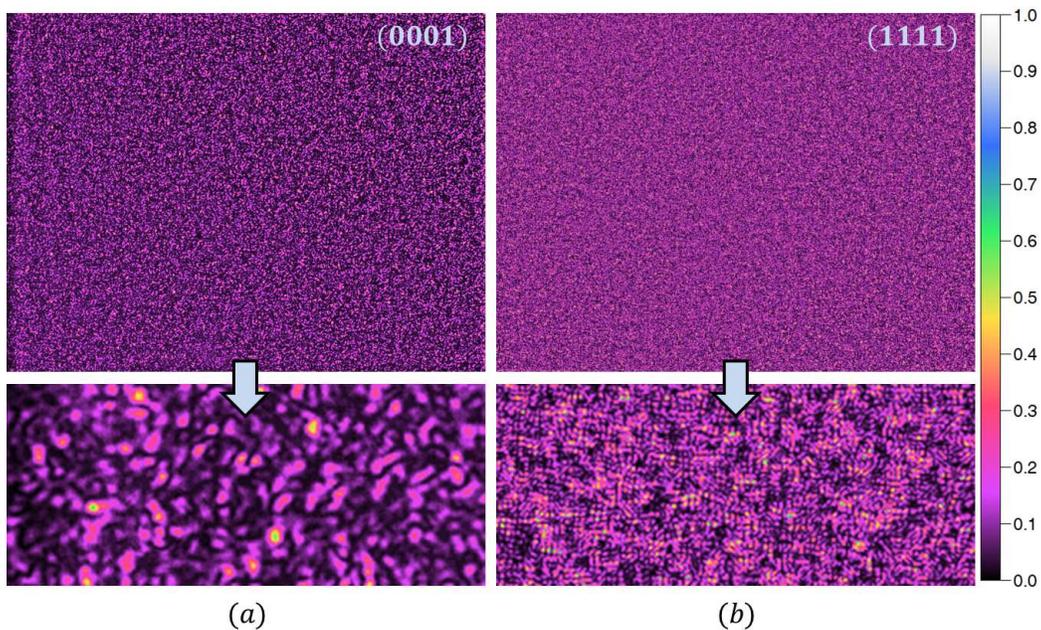}
\caption{\label{speckle}Images of the speckle fields (in false colours) acquired with the CCD for the two input combinations $(x_1,x_2,x_3,x_4)=(0,0,0,1)$ (a) and $(x_1,x_2,x_3,x_4)=(1,1,1,1)$ (b). The higher density of the speckles in (b), is due to the formation of interference fringes created by the superposition of the speckle fields produced with the four input laser beams. The bottom panels are a zoom of a portion of the top images.}
\end{figure}

In Fig. \ref{speckle} we show an example of a uniform speckle field obtained with two different input combinations: $(0,0,0,1)$ and $(1,1,1,1)$ respectively. Since the average intensity of the raw images of the speckle fields across the CCD is not uniform due to the different position of the lasers on the optical scatterer, it is useful, for statistical analysis, to make uniform the entire image. 

Therefore, the uniform speckle fields were numerically obtained by using the ratio $I(l,m)=I(l,m)_{CCD}/I(l,m)_{LPF}$, where $I(l,m)_{CCD}$ is the intensity of the speckle field acquired with the CCD, $I(l,m)_{LPF}$ is the same intensity filtered using a low-pass filter with a spatial frequency of 250 $\mu$m$^{-1}$, and $I(l,m)$ is the intensity of the corrected speckle field. Note that fields are characterized by the interference patterns, given by the superposition of the speckle fields from different inputs (Fig. \ref{speckle} (b)). The interference patterns have high contrast in a scale length typical of the transverse spatial coherence length $L_C \approx \lambda z/w$, where $w$ is the transverse beam size on the scatterer, $\lambda$ is the radiation wavelength and $z$ is the distance between the CCD and the scatterer. The random behavior of the observed intensity that characterizes the speckle fields is of fundamental importance for the receptron implementation, being a consequence of the large amount of the representing parameters in Eq. \ref{components}, providing to the network high variability and nonlinearity.

\subsection{Nonlinearity of the optical receptron}
\label{sec:nonlinearity}

\begin{figure}[http]
\includegraphics[scale=0.65]{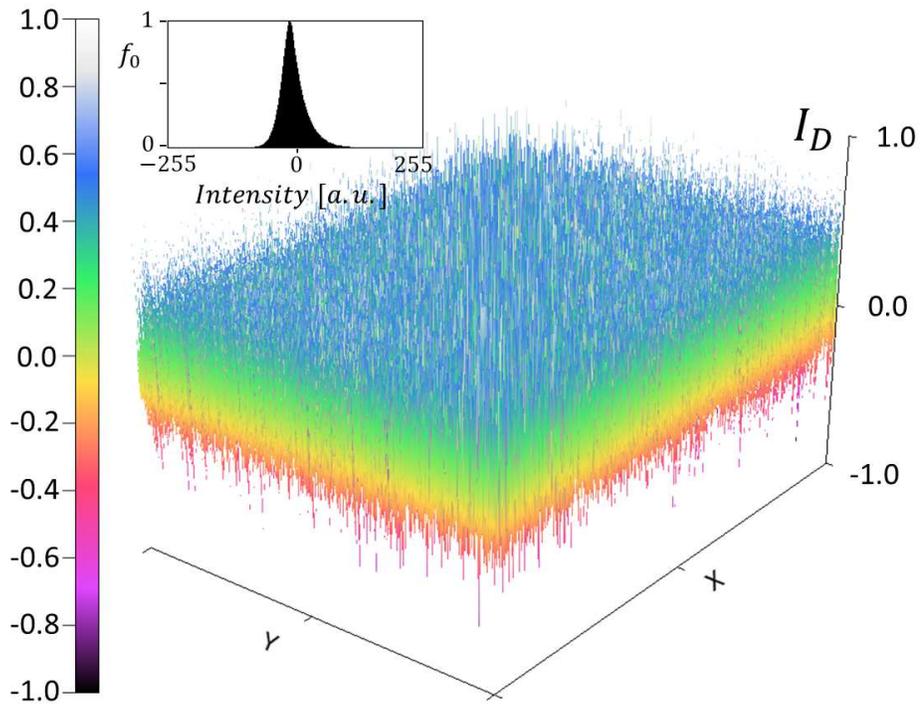}
\caption{\label{sovrapp}3D-view of the speckle field obtained with the difference $I_D$ normalized to the maximum intensity. The inset shows the intensity distribution of $I_D$ with the occurrence $f_0$ normalized to the maximum value. The RMS intensity ($I_{D(RMS)}=28$) is not in agreement with the superposition principle for linear systems ($I_{D(RMS)}=0$).}
\end{figure}

We checked the nonlinearity of the optical receptron by applying the superposition principle to the uncorrected intensity of the speckle field. Since $I_{CCD} (\vec{x})=S(\vec{x})$ (as discussed in section \ref{sec:optical_model}) we calculated $I_D=I_{CCD}(1,1,1,1)-[I_{CCD}(0,0,0,1)+I_{CCD}(0,0,1,0)+I_{CCD}(0,1,0,0)+I_{CCD}(1,0,0,0)]$. Nonlinearity of receptron has been represented in Fig. \ref{sovrapp} where we show $I_D$ and the relative distribution. We observe that the Root Mean Square (RMS) of $I_D$ ($\approx 28$, in a grey scale range between $255$ and $-255$) is comparable with those of the other speckle patterns. Moreover, we exclude that the observed fluctuations originate from noise, since noise has substantially lower RMS ($\sigma_n \approx 0.42$), concluding that speckle patterns behave with the expected nonlinearity with respect to the inputs. 

While the results shown in Fig. \ref{sovrapp} prove that the proposed system truly implements the receptron model, we are interested in a more general approach to define the non-linear properties. In the case of the optical receptron the tensor is reduced to a matrix, so we can use some of the standard measures for diagonality of a matrix. In particular, we can measure the relative importance of off-diagonal versus diagonal elements via the Frobenius norm \cite{Frobenius} for the optical receptron:

\begin{equation}\label{Frobenius}
\mu=\frac{||W-diag(W)||_F}{||W||_F}=\left(\frac{2\sum_{j<k}{w_{jk}^2}}{\sum_j{w_j^2}+2\sum_{j<k}{w_{jk}^2}}\right)^{1/2}
\end{equation}

where $W$ is the weights’ tensor and $||\cdot||_F$ is the Frobenius norm. The matrix elements  of the tensor have been extrapolated from data by inverting Eq. \ref{somma_primata}: the $2^n$ independent parameters for a given receptron are obtained from the system of $2^n$ equations by substituting into Eq. \ref{somma_primata} all the possible input combinations e.g. $w_3=I_{CCD}(0,0,1,0)-I_{CCD}(0,0,0,0)$, $w_{12}=I_{CCD}(1,1,0,0)-I_{CCD}(1,0,0,0)-I_{CCD}(0,1,0,0)-I_{CCD}(0,0,0,0)$. The coefficient $\mu$ lies between 0 and 1, where 0 indicates fully diagonal tensor and 1 a vanishing diagonal. 

\begin{figure}[http]
\includegraphics[scale=0.45]{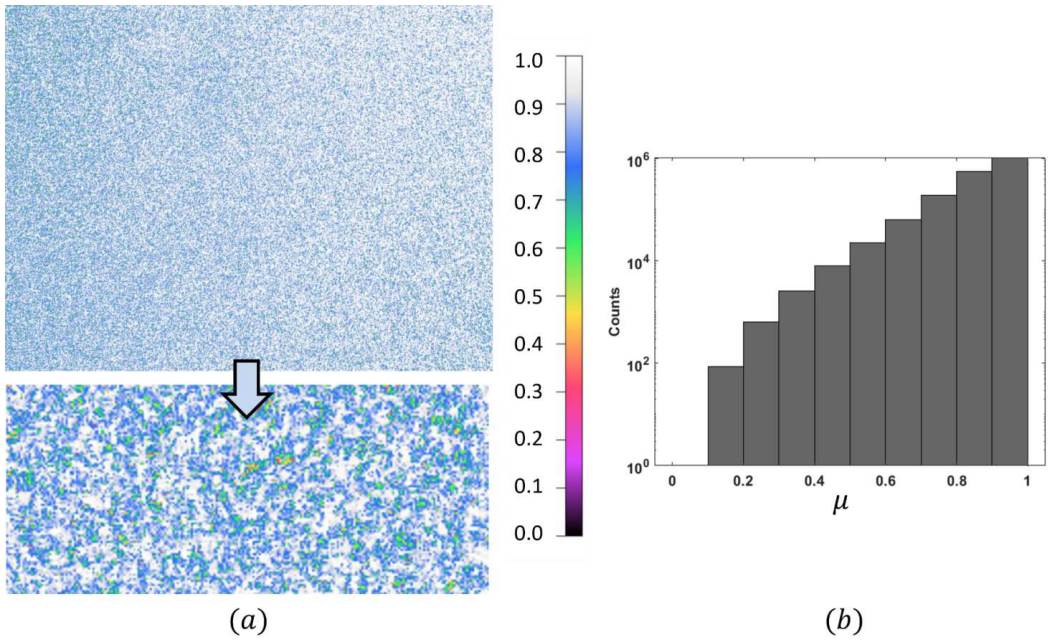}
\caption{\label{frobenius}(a) 2D-view of $\mu$ obtained for the same data shown in Fig. \ref{sovrapp} (bottom panel is a zoom of the top left corner). (b) Distribution of $\mu$ in semilog scale. Notice that the $\mu$ value of the largest number of pixels are concentrated close to unity, showing the expected highly nonlinear properties.}
\end{figure}

The high values achieved (see Fig. \ref{frobenius}) demonstrate the remarkable capabilities of the optical system, which reproduces almost all levels of linearity of a theoretical receptron.  Apart from confirming the result of Fig. \ref{sovrapp}, this example well demonstrates how the weights provide a powerful level of abstraction which is in turn an additional tool in the characterization of the system behavior.

\subsection{Implementation and classification of the Boolean functions}
\label{sec:Boolean_functions}

The implementation of different Boolean functions depending on the input combinations was verified experimentally with a set of 16 speckle fields as shown in Fig. \ref{setup} (c). These fields correspond to the digital combination of inputs ranging from $\vec{x}=(0,0,0,0)$ to $\vec{x}=(1,1,1,1)$ (or between 0 and 15 in the decimal base system). The gray scale images are converted in black and white images (Boolean fields), by applying a threshold $th$ in the range between 0 and 255 (8 bit grey scale range) in accordance with the thresholding process in Eq. \ref{condizioni}. Note that each pixel of the CCD (in grey scale) represents an independent function $S$, thus the intensity $I(l,m)=S(l,m)$ reproduces a collection of $l\times m$ non-linear functions, characterized by high statistical variability. Therefore the thresholding process generates $l\times m$ digital outputs $Y(l,m)$, each implementing its own Boolean function. 

Thanks to the high variability of the Boolean functions in the observation plane, the training of the network does not require an additional modification of the weights to solve a specific task. Network training is carried out for any function through the acquisition of 16 images, corresponding to the digital input combinations, and selecting the pixel (or pixels) that solve the target function regardless of the complexity of the function (see Fig. \ref{setup} (c)). Therefore, the training procedure consists in associating the position of the Boolean field (pixel coordinates) with the target function. The training procedure is then reduced to a well defined application and no time-consuming iterative algorithms are required to adjust the weights.

\begin{figure}[http]
\includegraphics[scale=0.6]{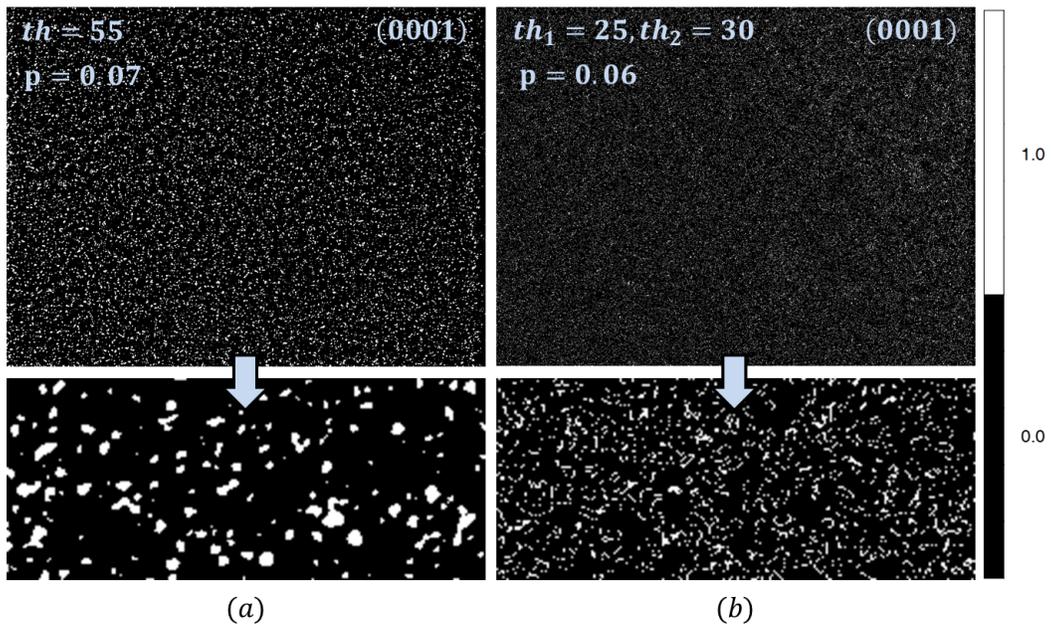}
\caption{\label{Boolean}Images of the Boolean fields generated with single threshold (a) and double threshold (b) for the input combination $(x_1,x_2,x_3,x_4)=(0,0,0,1)$. Despite the average success probability $p$ of (a) and (b) are similar, the double threshold drastically reduces the white speckles size. The bottom panels are a zoom of a portion of the top images.}
\end{figure}

\begin{figure}[http]
\includegraphics[scale=0.2]{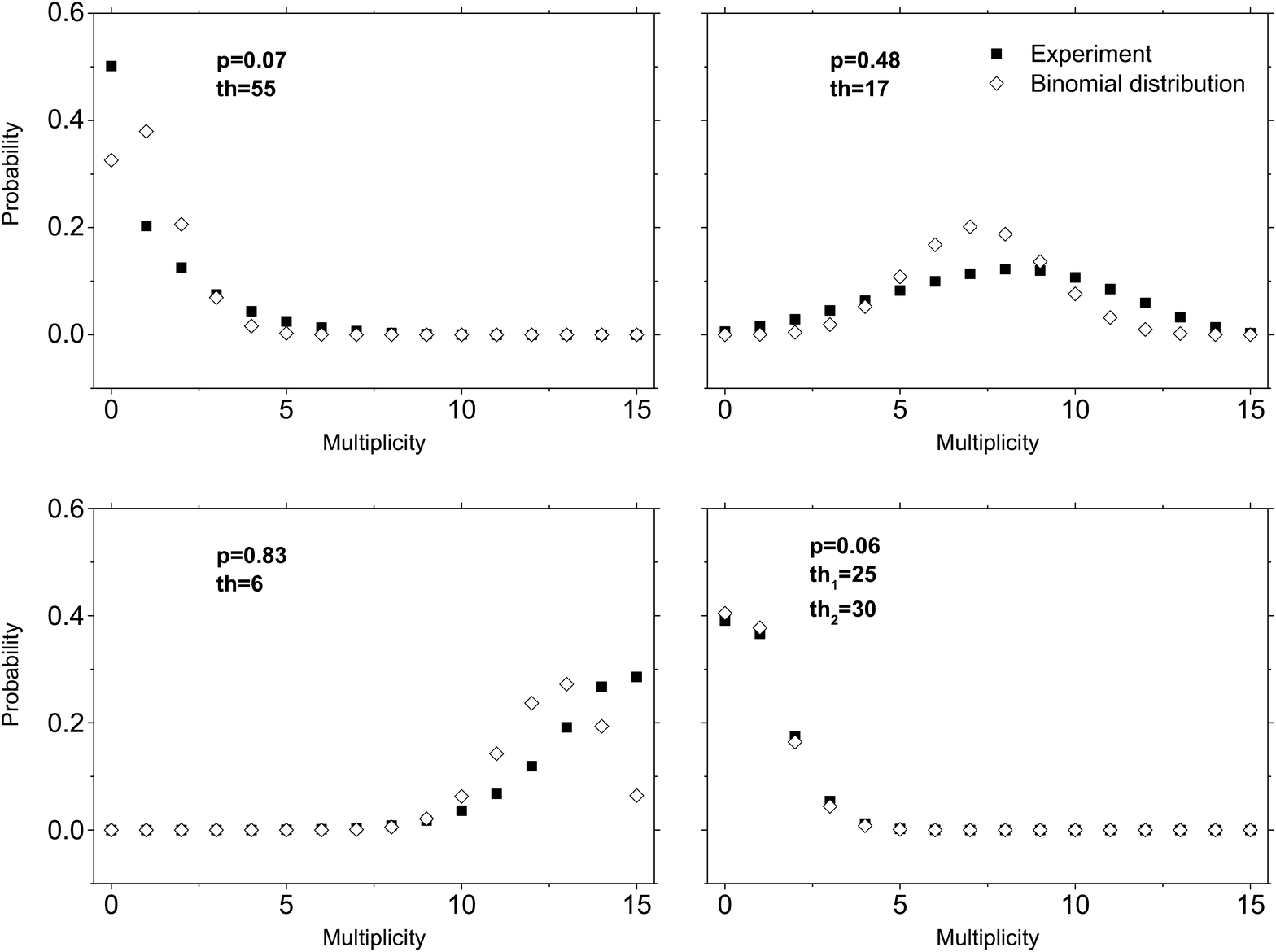}
\caption{\label{grafici}Experimental distributions of the functions classified with the multiplicity for different probabilities $p$. Data are compared with the theoretical binomial distributions. The distribution obtained with the double threshold (right-bottom) is in good agreement with the theory thanks to the minimization of pixel correlations generated by the speckle fields.}
\end{figure}

An example of Boolean field obtained for $th=55$ is shown in Fig. \ref{Boolean} (a). Notice that the white pixels are grouped in correspondence with the speckle areas providing a strongly correlated statistic. 
The statistical properties of the Boolean fields generated, deeply linked with their variability, can be controlled acting on the speckle spatial coherence length $L_C$ or on the threshold process.  The result can be evaluated with the multiplicity, i.e. the number of high states, of the Boolean function generated, which corresponds to the number of success events of a binomial statistical process made of $2^n$ trials.
The experimental distributions of the multiplicity of the output functions are shown in Fig. \ref{grafici}. Distributions are obtained on the sample of $1600 \times 1200$ pixels and for three different thresholds. Results can be compared with a binomial distribution, which is valid in the case of fully uncorrelated light intensities, having the success probability given by

\begin{equation}\label{probability1}
p=\frac{\sum_k{\int_{th}^{\infty}{\rho_k(I) dI}}}{2^n-1},
\end{equation}

where $\rho_k$ is the normalized intensity distribution ($\int_0^{\infty}{\rho_k(I) dI}=1$) of the speckle field with input combination $k$.

This shows that the peak of the experimental distributions can be shifted at low, medium and high multiplicity for the three probabilities $p=0,07$, $p=0,48$, $p=0,83$ acting on the threshold: this provides a method to promote the solution of functions with a given multiplicity. Note the deviations from the binomial model especially for the two middle scenarios: we interpret the difference as due to correlations generated by the coherence areas of the speckles, as also observed in the Boolean field of Fig. \ref{Boolean} (a).

To verify our interpretation, we generated a new set of 16 Boolean fields starting from the same experimental data by applying two thresholds $th_1=25$, $th_2=30$ (in a 8 bit grey scale range). The Boolean field was obtained by taking as "HIGH" state the pixel intensity of the corresponding speckle field in between the thresholds $th_1$, $th_2$ and by taking as "LOW" state the pixel intensity outside the two thresholds. Therefore, when the two thresholds are close each other, we expect a reduced speckles area composed of a few pixels, thus reducing the pixel correlation.
An example of Boolean field, obtained with the double threshold, is shown in Fig. \ref{Boolean} (b). Notice that the speckle areas, and hence the pixel correlations are drastically reduced. The corresponding distribution is shown in Fig. \ref{grafici} (right-bottom). The experimental distribution are now in good agreement with the corresponding theoretical binomial distribution with success probability,

\begin{equation}\label{probability2}
p=\frac{\sum_k{\int_{th_1}^{th_2}{\rho_k(I) dI}}}{2^n-1}.
\end{equation}

Summarizing, correlations can be used to change the shape of the multiplicity distribution, while the threshold gives control over the mean of such a distribution. 

\begin{figure}[http]
\includegraphics[scale=0.35]{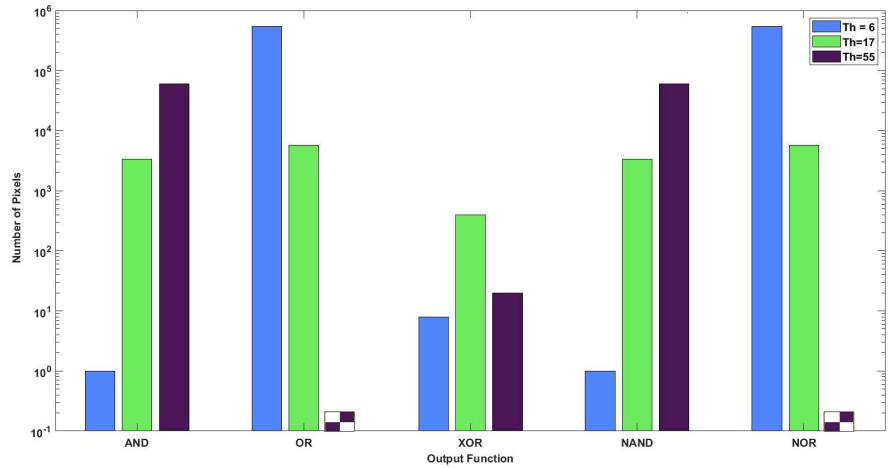}
\caption{\label{functions}Number of standard Boolean functions (AND, OR, XOR, NAND and NOR) solved by the network for different thresholds in logarithmic scale. Note that, due to complementary conditions shown in Eq. \ref{condizioni3}, the number of NOR and OR is the same, as well as for AND and NAND functions. The checkerboard bars for OR and NOR stand for zero pixels.}
\end{figure}

We have applied these tools to solve Boolean functions (Fig. \ref{functions}): in particular, we show the number of pixels that implement the functions $Y(x_1,x_2,x_3,x_4 )=x_1\cdot x_2 \cdot x_3 \cdot x_4$ (four-input AND gate), $Y(x_1,x_2,x_3,x_4 )=x_1+x_2+x_3+x_4$ (four-input OR gate) and $Y(x_1,x_2,x_3,x_4)=x_2 \oplus x_3$ as a function of the chosen threshold. The three functions have multiplicities $m=1$, $m=15$ and $m=8$, and are solved with higher efficiency (higher number of pixels) when the multiplicity is closer to the peak probability of the distributions shown in Fig. \ref{grafici}. It occurs for the three corresponding distributions having binomial success probabilities $p=0.07$, $p=0.83$, $p=0.48$ and obtained with the thresholds $th=55$, $th=6$, $th=17$, respectively. Similar considerations can be done to effectively solve the functions $Y(x_1,x_2,x_3,x_4)= \overline{x_1\cdot x_2 \cdot x_3 \cdot x_4}$ (NAND universal gate)  and  $Y(x_1,x_2,x_3,x_4)=\overline{x_1+x_2 +x_3 +x_4}$ (NOR universal gate). In fact, as shown in Fig. \ref{functions} we obtain the same number of pixels of the corresponding negated functions AND, OR by using the complementary conditions of Eq. \ref{condizioni} as

\begin{equation}\label{condizioni3}
Y(x_1,...,x_n) =
\left\{
\begin{array}{rl}
1 & \mbox{  } S\leq th  \\
0 & \mbox{  } S> th \:.
\end{array}
\right.
\end{equation}

These results confirm that by controlling the statistical properties of the resulting intensities we drastically increase the computing efficiency for a given Boolean algebraic problem.
Moreover, the nonlinear behavior experimentally verified in section \ref{sec:nonlinearity} makes it possible to solve non-linearly separable functions, as formally described in the example of Fig. \ref{xor}. The experimental result in Fig. \ref{functions} relative to XOR operator gives a direct proof of the receptron advantages with respect to the perceptron in terms of the type of generated functions.

\section{Conclusions}
\label{sec:conclusions}
We have developed a formal description of a device, the receptron, which generalizes the perceptron by considering that the input weights are not univocally related to a single input, hence they cannot be independently adjusted. This nonlinear characteristic is more similar to what observed in the interactions between synapses in neural dendritic trees \cite{Bicknell}.

An optical implementation of a receptron has been experimentally realized: the setup exploits simple optical elements to generate a large number of Boolean functions ($\approx 2 \cdot 10^6$) by using the high variability of the speckle fields. The complete set of functions has been classified with the multiplicity, in order to characterize their statistical properties. The role of the macroscopic parameters, such as the activation threshold, is essential to statistically control these properties. In fact, by changing the threshold the average multiplicity of the distributions can be changed in a controlled way. Furthermore, a control of the distribution shapes can be also managed by properly setting the double threshold in the thresholding process. These results show a method for solving (by shifting the peak probability or by changing the shape of the distribution) particular classes of Boolean functions and hence to statistically promote the solution of particular Boolean algebra problems. 

The support of a conventional computer was necessary in this proof-of-principle experiment to perform the thresholding process and the initial training procedure. Once these steps are realized, the optical system becomes independent on any conventional computer. For example, the speckle fields in correspondence of the CCD pixels, of one or more target functions, can be directly coupled with one or more photodiodes and amplifiers. In this way, the thresholding procedure can be realized with simple comparator circuits, by properly matching the computational and comparator thresholds. This method drastically increases the computing speed, being substantially limited only by the time responses of photodiodes, amplifiers and comparators with working frequencies above tens of GHz.

The optical implementation of the receptron scheme opens the way for the fabrication of a completely new class of optical devices for neuromorphic data processing based on a very simple hardware: a single receptron is already capable of solving non-linearly functions, therefore a network of receptrons could allow the combination of the outputs of several units to obtain the target function easier and faster. In stark contrast with a perceptron network, an optical receptron nework would not intrinsically require more energy than its single-element counterpart. In other words, a receptron can already generate all functions in principle, but the search for a desired function could take too long, due to the exponential increase in the number of possibilities as a function of the inputs. A simple combination of receptrons could instead simplify the search. Since the number of parameters is very large compared to a perceptron, one does not need a fine control of each parameter to obtain a given result, since it can be obtained with different configurations of the same parameters. These aspects are of fundamental interest in view of the fabrication and scale-up of optical receptron networks for very complex data processing tasks.



 \bibliographystyle{elsarticle-num} 


\end{document}